\documentclass[twocolumn,aps,tightenlines]{revtex4}
\usepackage{lmodern}
\usepackage{lmodern}
\usepackage[latin9]{inputenc}
\usepackage{amsmath}
\usepackage{amssymb}
\usepackage{graphicx}
\usepackage{esint}

\makeatletter

\newcommand*\LyXThinSpace{\,\hspace{0pt}}

\@ifundefined{textcolor}{}
{%
 \definecolor{BLACK}{gray}{0}
 \definecolor{WHITE}{gray}{1}
 \definecolor{RED}{rgb}{1,0,0}
 \definecolor{GREEN}{rgb}{0,1,0}
 \definecolor{BLUE}{rgb}{0,0,1}
 \definecolor{CYAN}{cmyk}{1,0,0,0}
 \definecolor{MAGENTA}{cmyk}{0,1,0,0}
 \definecolor{YELLOW}{cmyk}{0,0,1,0}
}

\makeatother

\begin{document}

\title{The origin of nematic order in FeSe}

\author{Andrey V. Chubukov$^{1}$, Rafael M. Fernandes$^{1}$, and Joerg
Schmalian$^{2}$}

\affiliation{$^{1}$ School of Physics and Astronomy, University of Minnesota,
Minneapolis, MN 55455, USA \\
 $^{2}$ Institite for Theory of Condensed Matter and Institute for
Solid State Physics, Karlsruhe Institute of Technology, Karlsruhe,
Germany 76131}
\begin{abstract}
The origin of the $90$ K nematic transition in the chalcogenide FeSe,
which displays no magnetic order down to $T=0$, remains a major puzzle
for a unifying theory for the iron-based superconductors. We analyze
this problem in light of recent experimental data which reveal very
small Fermi pockets in this material. We show that the smallness of
the Fermi energy leads to a near-degeneracy between magnetic fluctuations
and fluctuations in the charge-current density-wave channel. While
the two fluctuation modes cooperate to promote the same  preemptive Ising-nematic
order, they compete for primary order. We argue that this explains
why in FeSe the nematic order emerges when the magnetic correlation
length is smaller than in other Fe-based materials, and why no magnetism
is observed. We discuss how pressure lifts this near-degeneracy, resulting
in a non-monotonic dependence of the nematic transition with pressure,
in agreement with experiments.
\end{abstract}
\maketitle
Nematic order in Fe-pnictides and Fe-chalcogenides develops at a temperature
$T_{s}$ that is larger than the magnetic transition (for reviews,
see \cite{review}). It spontaneously breaks the tetragonal $C_{4}$
lattice symmetry down to orthorhombic $C_{2}$. The origin of this
symmetry breaking is currently one of the most intensely debated issues
of the Fe-based superconducting materials \cite{Fernandes14}. In
the Fe-pnictides, nematic order occurs reasonably close to the instability
towards stripe magnetic order at the Neel temperature $T_{N}$. Because
the stripe order breaks $Z_{2}$ tetragonal symmetry ($C_{4}\to C_{2}$)
in addition to the $O(3)$ spin-rotational symmetry and because $T_{s}$
and $T_{N}$ show similar doping dependencies, it seems reasonable
to associate the nematic order with magnetism~\cite{Fernandes14}.
Indeed, several groups have
 argued~\cite{Xu08,Fang08,Si11,igor_m,Batista11,Lorenzana11,Brydon11,Fernandes12,Dagotto13,Yamase15} that
  magnetic fluctuations split the mean-field
stripe magnetic transition into two separate $O(3)$ and $Z_{2}$
transitions. The discrete $Z_{2}$ symmetry is broken first at $T_{s}>T_{N}$,
resulting in an intermediate phase, dubbed Ising-nematic, where long-range
magnetic order is absent but the $C_{4}$ lattice symmetry is broken
down to $C_{2}$. Such $Z_{2}$ order triggers orbital and structural
order as all three
 break the same $C_{4}$ symmetry.

The magnetic scenario for nematicity in Fe-pnictides is supported
by a variety of experimental observations, such as the doping dependencies
of $T_{N}$ and $T_{s}$ \cite{Fernandes12}, the scaling between
the shear modulus and the spin-lattice relaxation rate \cite{Fernandes13},
and the sign-change of the in-plane resistivity anisotropy between
electron-doped and hole-doped Fe-pnictides\cite{Blomberg13}. This
scenario, however, has been challenged for the Fe-chalcogenide FeSe.
This material displays a nematic transition at $T_{s}\approx90K$.
The properties of the nematic phase in FeSe resemble those in Fe-pnictides:
similar softening of the shear modulus \cite{Meingast_FeSe}, similar
orthorhombic distortion and orbital order \cite{Nakayama_ARPES_14,Ding_ARPES_15,ZXShen_ARPES_15},
and similar behavior of the resistivity anisotropy upon applied strain
\cite{Coldea15}. Furthermore, neutron scattering experiment shows
that spin fluctuations are peaked at the same ordering vectors as
in the Fe-pnictides \cite{INS_FeSe_1,INS_FeSe_2}. Yet, in distinction
to Fe-pnictides, no magnetic order has thus far been observed in FeSe
in the absence of external pressure \cite{McQueen09,Imai09}. Moreover,
NMR measurements were interpreted as evidence that the magnetic correlation
length $\xi$ remains small at $T_{s}$ \cite{Buchner_FeSe,Meingast_FeSe}.
Although in the Ising-nematic scenario $\xi$ \textit{does not} \emph{have
to be large} at $T_{s}$, this seems to be the case for all Fe-pnictides.

Given these difficulties with the Ising-nematic scenario,
  spontaneous
   orbital
order has been invoked to explain the nematic state in FeSe \cite{Buchner_FeSe,Meingast_FeSe}.
However, at present, no microscopic theory exists where orbital order
appears spontaneously instead of being induced by magnetism~\cite{w_ku10,devereaux10,Phillips10,Phillips12,Kontani12}.
Alternative scenarios for magnetically-driven nematicity in FeSe have
also been proposed, involving the formation of a quantum paramagnet
\cite{Kivelson_Lee_15}, the onset of spin quadrupolar order \cite{Si15},
and strong frustration of the magnetic fluctuations \cite{Mazin15}.
Yet, the issue of why FeSe does not fit into a ``universal'' theory
for the iron-based superconductors still persists.

In this communication, we present an extension of the spin-nematic
scenario which explicitly builds on a unique property of the electronic
structure of FeSe, namely, the fact that the Fermi energy $E_{F}$
in this material is
 small -- only a few meV, as seen by ARPES
and dH-vA experiments \cite{Coldea15,FeSe_dHvA}. For a system with
a small $E_{F}$, earlier renormalization-group (RG) calculations
have shown that there are two density-wave channels whose fluctuations
are strong at momenta $(0,\pi)/(\pi,0)$: a spin density-wave (SDW)
channel and a charge-current density-wave (CDW) channel (a CDW with
imaginary order parameter, which we denote as iCDW \cite{zlatko}).
The relative strength between the two depends on the sign of the inter-pocket
exchange interaction ($u_{2}$ in our notations below). For repulsive
$u_{2}$, the coupling in the SDW channel is larger, while for attractive
$u_{2}$ the coupling in the iCDW channel is larger. In both cases,
however, the RG calculations show that the coupling in the subleading
channel approaches the one in the leading channel at small energies.
The RG process stops at $E_{F}$, implying that if $E_{F}$ is larger
than the highest instability temperature ($T_{s}$ for FeSe) the subleading
channel is not a strong competitor and for all practical purposes
can be neglected. However, if
 $E_{F} \sim T_{s}$,
  as in FeSe, the couplings
in the two channels become
 degenerate within the RG. The degeneracy
implies that the order parameter manifold increases from $O(3)\times Z_{2}$,
for the three-component SDW, or from $Z_{2}\times Z_{2}$, for the
one-component iCDW, to a larger $O(4)\times Z_{2}$. In all cases,
the $Z_{2}$ part of the manifold corresponds to selecting either
$(0,\pi)$ or $(\pi,0)$ for the density-wave ordering vector. While
in both $O(3)\times Z_{2}$ and $O(4)\times Z_{2}$ models the $Z_{2}$
symmetry can be broken before the continuous one, in the latter this
happens at a significantly smaller correlation length. As a result,
at small $E_{F}$, the nematic order emerges while magnetic fluctuations
are still weak. Furthermore, the SDW transition temperature $T_{N}$
in the $O(4)$ model is additionally suppressed due to the competition
with iCDW. We argue that these features explain the
 properties
of the nematic state in FeSe, including
non-monotonic
pressure dependence of $T_{s}$~\cite{FeSe_pressure1,FeSe_pressure2}.

\textit{The model.}~~~We consider a quasi-2D itinerant band model
with two hole pockets
at the $\Gamma$ point and two electron
pockets
at $(0,\pi)$ and $(\pi,0)$ in the 1-Fe Brillouin
zone \cite{Eremin10,Fernandes12}. This
model can be obtained
from an underlying 5-orbital model with Hubbard and Hund interactions
and hopping between the Fe $3d$ orbitals
\cite{GraserSDDeg,comm}.

\begin{figure}
\includegraphics[width=0.8\columnwidth]{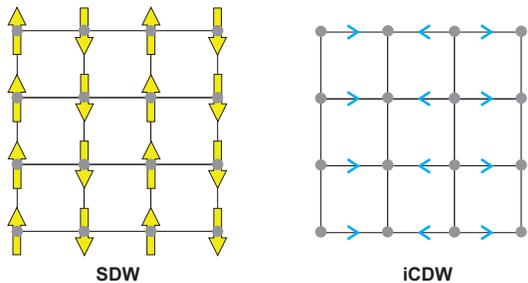} \protect\protect\protect\protect\protect\caption{Schematic representation of the SDW and iCDW ordered states with ordering
vector $\left(\pi,0\right)$. In the latter, the arrows represent
charge currents along the bonds, while in the former, they represent
spins on the sites. While fluctuations of both channels support nematicity,
they compete for long-range magnetic and charge order. \label{fig_SDW_iCDW}}
\end{figure}

The quadratic part of the Hamiltonian in the band basis describes
the dispersion of the low-energy fermions, and the information about
the orbital content along the Fermi pockets is passed onto inter-pocket
and intra-pocket interactions, which are the Hubbard and Hund terms
dressed by the matrix elements arising from the change from the orbital
to the band basis \cite{Valenzuela14}. The angular dependence of
the matrix elements leads to angle-dependent interactions. The three
interactions relevant for Ising-nematic order are the inter-pocket
density-density interaction $u_{1}$, the exchange interaction $u_{2}$,
and the pair-hopping interaction $u_{3}$ \cite{Chubukov_RG}. To
simplify the analysis, we follow earlier works \cite{Fernandes12}
and analyze the Ising-nematic order within an RG procedure that (i)
approximates these three interactions as angle-independent and (ii)
restricts the analysis to one hole pocket. The extension to two pockets
and angle-dependent interactions makes the calculations more involved
but does not modify the RG equations in any substantial way and, moreover,
leaves them intact if we treat the hole pockets as circular and neglect
the $d_{xy}$ orbital component on the electron pockets.

We label the fermions near the hole pocket as $c_{\mathbf{k}}$ and
the fermions near the electron pockets as $f_{1,\mathbf{k}}$ and
$f_{2,\mathbf{k}}$. 
 The $O(3)$ magnetic
order parameter is given by
\begin{equation}
{\bf M}_{j}=\frac{1}{N}\sum_{\mathbf{k}\alpha\beta}\left(c_{\mathbf{k},\alpha}^{\dagger}{\bf \boldsymbol{\sigma}}_{\alpha\beta}f_{j,\mathbf{k}+\mathbf{Q}_{j},\beta}+h.c\right),
\end{equation}
whereas 
the $Z_{2}$ iCDW order parameter is
\begin{equation}
\Phi_{j}=\frac{i}{N}\sum_{\mathbf{k}\alpha}\left(c_{\mathbf{k},\alpha}^{\dagger}f_{j,\mathbf{k}+\mathbf{Q}_{j},\alpha}-h.c.\right),
\end{equation}
with $j=1,2$ corresponding to the two possible ordering vectors $\mathbf{Q}_{1}=(\pi,0)$
and $\mathbf{Q}_{2}=(0,\pi)$. We show these two ordered states in
Fig. \ref{fig_SDW_iCDW}.

\textit{$O(4)$ Ising-nematic action.}~~~ In the Ising-nematic
scenario, the $C_{4}\to C_{2}$ symmetry breaking implies the appearance
of a composite order, quadratic in the density-wave order parameters
${\bf M}_{j}$ and $\Phi_{j}$. To analyze this scenario, we need
to know the flow of the couplings that drive SDW order, $\Gamma_{\mathrm{sdw}}=u_{1}+u_{3}$,
and iCDW order, $\Gamma_{\mathrm{icdw}}=u_{1}+u_{3}-2u_{2}$ (Ref.
\cite{Chubukov_RG}). The bare coupling $\Gamma_{\mathrm{sdw}}>\Gamma_{\mathrm{icdw}}$
when $u_{2}>0$ and $\Gamma_{\mathrm{icdw}}>\Gamma_{\mathrm{sdw}}$
when $u_{2}<0$. As one integrates out the high-energy degrees of
freedom via an RG procedure, the ratio $u_{2}/(u_{1}+u_{3})$ decreases
as the system flows to lower energies (or temperatures) and approaches
zero at the energy/temperature scale in which the system develops
SDW/iCDW order. This holds, however, only if this scale is larger
than $E_{F}$. If $E_{F}$ is larger, the RG flow stops at $E_{F}$
and the system develops an instability only in the channel with the
largest bare coupling.

\begin{figure}
\includegraphics[width=0.8\columnwidth]{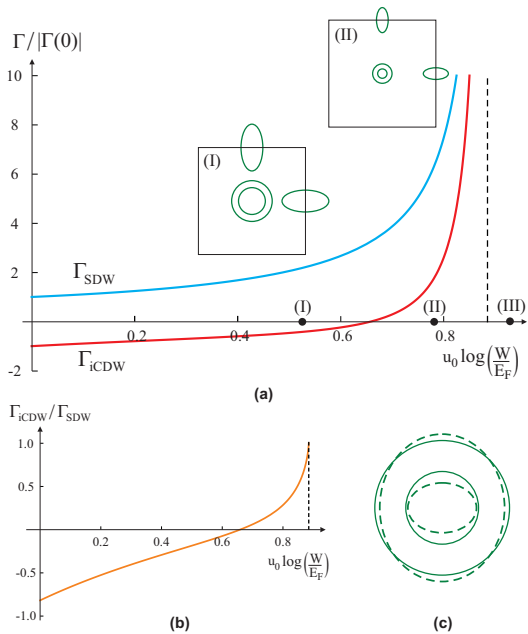} \protect\protect\protect\protect\protect\caption{(a) RG flow of the SDW and iCDW interactions $\Gamma_{\mathrm{sdw}}$
(red curve) and $\Gamma_{\mathrm{icdw}}$ (blue curve) as function
of decreasing energy $E$. $W$ is the bandwidth, $u_{0}=u_{1}\left(0\right)=u_{2}\left(0\right)=10u_{3}\left(0\right)$
is the bare interaction parameter, and the dashed line is the energy
in which the two degenerate instabilities occur. The RG flow stops
at the Fermi energy $E_{F}$: if $E_{F}$ is large (case I, Fe-pnictides),
only SDW fluctuations are relevant, whereas if $E_{F}$ is small (cases
II/III, FeSe), both SDW and iCDW fluctuations are important. The insets
show schematically the Fermi pockets in each case. (b) Ratio $\Gamma_{\mathrm{icdw}}/\Gamma_{\mathrm{sdw}}$
along the RG flow. (c) Electronic manifestation of the Ising-nematic
order on the hole pockets. There is a $\cos2\theta$ distortion, with
opposite signs for the two pockets, and an overall shift of the chemical
potential. \label{fig_RG_flow}}
\end{figure}

To illustrate our point, we plot in Fig. \ref{fig_RG_flow}(a)-(b)
the RG flow of $\Gamma_{\mathrm{icdw}}$ and $\Gamma_{\mathrm{sdw}}$
for a particular set of bare couplings $u_{1}\left(0\right)=u_{2}\left(0\right)=10u_{3}\left(0\right)$,
chosen deliberately to give a negative bare $\Gamma_{\mathrm{icdw}}$.
Under the RG flow, $\Gamma_{\mathrm{icdw}}$ becomes positive and
approaches $\Gamma_{\mathrm{sdw}}$ at the scale where the couplings
diverge and the system develops a density-wave order. The Fermi energy
$E_{F}$ sets the scale at which the RG flow stops. In case I (large
$E_{F}$), the RG stops when $\Gamma_{\mathrm{icdw}}$ is still small.
In case II (smaller $E_{F}$), the RG stop when $\Gamma_{\mathrm{icdw}}$
is comparable to $\Gamma_{\mathrm{sdw}}$, and in case III (even smaller
$E_{F}$), the RG flow reaches the $O(4)$ fixed point already at
energies larger than $E_{F}$. We associate case I in Fig. \ref{fig_RG_flow}
with Fe-pnictides, and cases II/III with FeSe based on the values
of $E_{F}$ obtained by ARPES and quantum oscillations \cite{FeSe_dHvA,Coldea15}.

We next take the RG results as input and analyze the emergence of
a nematic order which spontaneously breaks the symmetry between momenta
${\bf Q}_{1}$ and ${\bf Q}_{2}$ without breaking any other symmetry.
The analysis follows the same steps as for pure SDW order \cite{Fernandes12}:
we introduce ${\bf M}_{j}$ and $\Phi_{j}$ ($j=1,2$) as Hubbard-Stratonovich
fields which decouple the four-fermion interaction terms, integrate
over the fermions, and obtain the effective action in terms of ${\bf M}_{j}$
and $\Phi_{j}$:

\begin{eqnarray}
S_{\mathrm{eff}} & = & \int_{qj}\left(\chi_{s,q}^{-1}\mathbf{M}_{j}^{2}+\chi_{c,q}^{-1}\Phi_{j}^{2}\right)+\frac{u}{2}\int_{xj}\left(\mathbf{M}_{j}^{2}+\Phi_{j}^{2}\right)^{2}\nonumber \\
 & - & \frac{g}{2}\int_{x}\left[\left(\mathbf{M}_{1}^{2}+\Phi_{1}^{2}\right)-\left(\mathbf{M}_{2}^{2}+\Phi_{2}^{2}\right)\right]^{2}\label{S_eff}
\end{eqnarray}
where $\chi_{s,q}^{-1}=\Gamma_{\mathrm{sdw}}^{-1}-\Pi_{q}$ and $\chi_{c,q}^{-1}=\Gamma_{\mathrm{icdw}}^{-1}-\Pi_{q}$
with $\Pi_{q}=\int_{k}G_{c,k+q}\left(G_{f_{1},k}+G_{f_{2},k}\right)$.
Note, the only asymmetry between the two order parameters is due to
the interactions $\Gamma_{{\rm sdw}}$ and $\Gamma_{{\rm icdw}}$,
respectively. Near ${\bf Q}_{j}$, we can expand $\chi_{s(c),q}^{-1}\approx r_{0,s(c)}+\alpha({\bf q}-{\bf Q}_{j})^{2}$,
where $r_{0,s(c)}$ measures the distance to the SDW (iCDW) mean-field
instability and $\alpha\sim\mathcal{O}(1)$. The input from the RG
analysis is that $r_{0,s}$ and $r_{0,c}$ are close to each other.
The quartic coefficients are given by $(u,g)=\pm\frac{1}{2}\int_{k}G_{c,k}^{2}\left(G_{f_{1},k}\pm G_{f_{2},k}\right)^{2}$.
At $\Gamma_{\mathrm{sdw}}=\Gamma_{\mathrm{icdw}}$, the action depends
on ${\bf M}$ and $\Phi$ only via the combination ${\bf M}^{2}+\Phi^{2}$,
and the order parameter manifold is $O(4)\times Z_{2}$. Evaluating
the integrals
 at $E_F \sim T_s$,
  we find $u>0$ and $g>0$, what implies that long-range
order selects either $j=1$ or $j=2$, but not both, i.e. it breaks
both $O(4)$ and $Z_{2}$ symmetries.

Within a mean-field approximation, $O(4)$ and $Z_{2}$ are broken
at the same temperature. Beyond mean-field, the $Z_{2}$ symmetry
is broken first, and \textit{both} ${\bf M}$ and $\Phi$ contribute
to it, even if $\Gamma_{\mathrm{sdw}}\neq\Gamma_{\mathrm{icdw}}$.
To see this, we treat ${\bf M}$ and $\Phi$ as fluctuating fields,
introduce the composite fields $\psi=u\left(\mathbf{M}_{x}^{2}+\Phi_{x}^{2}+\mathbf{M}_{y}^{2}+\Phi_{y}^{2}\right)$
and $\varphi=g\left(\mathbf{M}_{x}^{2}+\Phi_{x}^{2}-\mathbf{M}_{y}^{2}+\Phi_{y}^{2}\right)$
to decouple the quartic terms, integrate over the primary fields ${\bf M}$
and $\Phi$ and obtain the action in terms of $\psi$ and $\varphi$:
\begin{eqnarray}
S_{\mathrm{eff}}\left[\varphi,\psi\right] & = & \frac{\varphi^{2}}{2g}-\frac{\psi^{2}}{2u}+\frac{3}{2}\int_{q}\ln\left[\left(\chi_{s}^{-1}+\psi\right)^{2}-\varphi^{2}\right]\nonumber \\
 & + & \frac{1}{2}\int_{q}\ln\left[\left(\chi_{c}^{-1}+\psi\right)^{2}-\varphi^{2}\right]\label{e_1}
\end{eqnarray}
The field $\psi$ has a non-zero expectation value $\left\langle \psi\right\rangle \neq0$
at any tenperature as it does not break any symmetry, but only renormalizes
the correlation lengths of the primary fields ${\bf M}$ and $\Phi$
to $\xi_{s(c)}^{-2}=r_{0,s(c)}+\left\langle \psi\right\rangle $.
A non-zero $\left\langle \varphi\right\rangle $, on the other hand,
breaks the tetragonal $C_{4}$ symmetry. If this happens before the
susceptibilities of the primary fields soften at ${\bf Q}_{j}$, then
the $Z_{2}$ rotational symmetry breaks prior to other symmetry breakings.
We emphasize that the nematic order parameter $\varphi$ involves
the combination ${\bf M}^{2}+\Phi^{2}$, hence one cannot separate
SDW induced and iCDW induced nematic order, even when $\chi_{s}$
and $\chi_{c}$ are not equivalent.

We solve the action in (\ref{e_1}) within the saddle-point approximation,
similarly to what was done in Refs.\cite{Fernandes12}. We find that
at $\xi_{s},\xi_{c}\approx\xi$, a non-zero nematic order parameter
$\left\langle \varphi\right\rangle \neq0$ emerges when the correlation
length $
\xi^{2}=
\pi/g$, or, to logarithmic accuracy in $g\ll1$, at $T_{s}=
2\pi\rho_{s}/|\log g|$, where $\rho_{s}$ is the stiffness of the $O(4)$ non-linear $\sigma$
model associated with Eq. (\ref{S_eff}). It is instructive to compare
this result with the case where only $O(3)$ SDW fluctuations are
present. In that case, the nematic order emerges when $3\xi_{O(3)}^{2}=4\pi/g_{O(3)}$,
and the transition temperature is $T_{s}=
2\pi\rho_{s}/|\log\sqrt{g_{O(3)}}|$, where $g_{O(3)}$ is the coupling in the SDW $O(3)$ model. As a
result, to obtain the same $T_{s}$, one needs a much smaller coupling
constant $g_{O(3)}\sim g_{O(4)}^{2}$. Consequently, at $T=T_{s}$,
the correlation length $\xi_{O(4)}$ in the $O(4)$ case is proportional
to $\xi_{O(4)}\sim\sqrt{\xi_{O(3)}}$, i.e. it is much smaller than
it would be if nematicity was driven solely by SDW fluctuations. This
is consistent with NMR \cite{Buchner_FeSe,Meingast_FeSe} and neutron
scattering data \cite{INS_FeSe_2} in the paramagnetic phase of FeSe,
which point to the presence of SDW fluctuations, albeit weaker than
in the Fe-pnictide compounds. The rapid increase of the correlation
lengths below $T_{s}$, obeying $\xi_{s,c}^{-2}=\xi_{s,c}^{-2}\left(T_{s}\right)-\left\langle \varphi\right\rangle $,
is also consistent with the increase of $1/T_{1}T$ and the inelastic
neutron signal\cite{Buchner_FeSe,Meingast_FeSe,INS_FeSe_2}.

The $O(4)$ Ising-nematic scenario also addresses why no magnetic
order appears down to the lowest temperatures. The SDW and iCDW orders
compete via the bi-quadratic term $\left(u-g\right)\mathbf{M}_{j}^{2}\Phi_{j}^{2}$
in the low-energy action of Eq. (\ref{S_eff}). As a result, for $u_{2}>0$,
fluctuations of the sub-leading iCDW channel suppress the transition
temperature of the leading SDW channel. Such a suppression is the
largest when the difference between the coupling constants $\left|\Gamma_{\mathrm{sdw}}-\Gamma_{\mathrm{icdw}}\right|$
is the smallest, which happens when the system flows towards $O(4)$
symmetry within RG, i.e. when $E_{F}$ is small, such as in FeSe.

\begin{figure}
\centering{}\includegraphics[width=0.8\columnwidth]{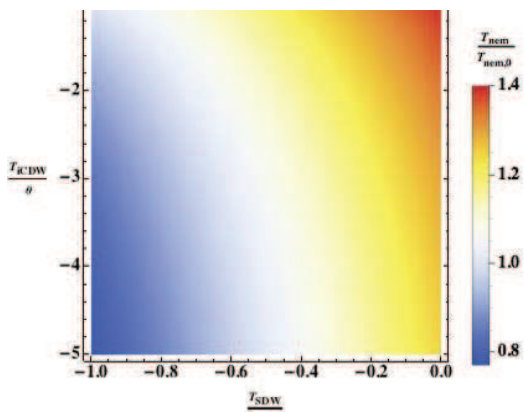}
\protect\protect\protect\protect\caption{Density plot of the nematic transition $T_{\mathrm{nem}}$ as function
of the bare iCDW and SDW transitions $T_{\mathrm{iCDW}}$ and $T_{\mathrm{SDW}}$.
To mimic the effect of pressure, they start at the same negative value
$-\theta$ at zero pressure, for which the nematic transition temperature
is $T_{\mathrm{nem,0}}$, and then vary in opposite ways upon increasing
pressure, \textit{\emph{$T_{\mathrm{iCDW}}<-\theta$ and $T_{\mathrm{SDW}}>-\theta$.}}
\label{Fe_pressure}}
\end{figure}

\textit{Experimental signatures.}~~~We know discuss the experimental
consequences of the Ising-nematic order. The breaking of the $Z_{2}$
symmetry between the $j=1$ and $j=2$ components of the $O(4)$ field
implies the breaking of $C_{4}$ lattice rotational symmetry down
to $C_{2}$. This instantaneously triggers structural order due to
the coupling to lattice. To investigate how $Z_{2}$ order affect
the electronic states, we return to the original four-pocket model
(with fermions near the two hole pockets described by the operators
$c_{1,\mathbf{k}}$ and $c_{2,\mathbf{k}}$) and include the explicit
angle-dependence introduced by the matrix elements for the transformation
between orbital and band basis. This transformation has the particularly
simple form $c_{1,\mathbf{k}}=d_{xz}\cos{\theta_{\mathbf{k}}}-d_{yz}\sin{\theta_{\mathbf{k}}}$,
$c_{2,\mathbf{k}}=d_{xz}\sin{\theta_{\mathbf{k}}}+d_{yz}\cos{\theta_{\mathbf{k}}}$
if one considers circular hole pockets and neglects the $d_{xy}$
orbital component on the electron pockets
~\cite{Vafek_Fernandes}

The feedback effect of the Ising-nematic order on the fermions takes
place via the self-energy corrections involving the unequal susceptibilities
of the primary SDW and iCDW fields at momenta $\mathbf{Q}_{1}$ and
$\mathbf{Q}_{2}$. These corrections not only shift the chemical potentials
of the $f_{1}$ and $f_{2}$ electron pockets in opposite directions
$\left\langle f_{1,\mathbf{k}}^{\dagger}f_{1,\mathbf{k}}\right\rangle -\left\langle f_{2,\mathbf{k}}^{\dagger}f_{2,\mathbf{k}}\right\rangle \propto\left\langle \varphi\right\rangle $,
but also give rise to a $d$-wave like distortion of the $c_{1}$
and $c_{2}$ hole pockets: $\left\langle c_{1,\mathbf{k}}^{\dagger}c_{1,\mathbf{k}}\right\rangle -\left\langle c_{2,\mathbf{k}}^{\dagger}c_{2,\mathbf{k}}\right\rangle \propto\left\langle \varphi\right\rangle \cos2\theta_{\mathbf{k}}$
(see Fig.\ref{fig_RG_flow}b). In the orbital basis, the latter corresponds
to ferro-orbital order $\left\langle d_{xz}^{\dagger}d_{xz}\right\rangle -\left\langle d_{yz}^{\dagger}d_{yz}\right\rangle \propto\left\langle \varphi\right\rangle $
\cite{Vafek_Fernandes}. Note that besides the changes in the dispersions
proportional to $\left\langle \varphi\right\rangle $, there
 is an overall shift of the chemical potential,
 symmetric for the two electron and the two hole pockets.

The behavior of hole
 pockets in the Ising-nematic scenario
is consistent with the existing ARPES data that show a $d$-wave type
elongation of one of the hole pockets, whereas the other hole pocket
sinks below the Fermi level \cite{Coldea15}. The behavior of the
electron pockets in the 2-Fe Brillouin zone is also consistent with
the splitting of the chemical potentials of the $f_{1}$ and $f_{2}$
fermions.

We also investigate how pressure affects the nematic transition temperature
$T_{s}$. Within our approach, $T_{s}$ is defined by the condition
$3\xi_{s}^{2}+\xi_{c}^{2}=4\pi/g$. Upon pressure, the Fermi pockets
become bigger, and the Fermi energy increases. As a result
 iCDW becomes less competitive and $\xi_{c}$ decreases,
while $\xi_{s}$ increases. The combination of these two opposite
tendencies in general gives rise to a non-monotonic behavior of $T_{s}$.
This is illustrated in Fig. \ref{Fe_pressure} using a simple modeling
 in which $\xi_{j}^{-2}\approx T-T_{j}$, with $T_{j}$
denoting the bare transition temperatures for SDW and iCDW
 (see caption).

Note that in our analysis so far we considered $u_{2}\left(0\right)>0$.
If on the other hand this interaction is attractive, $u_{2}(0)<0$,
the iCDW phase is the leading instability, and the ground state manifold
is $Z_{2}\times Z_{2}$. In this case, the nematic and iCDW transitions
are expected to be simultaneous \cite{Fernandes12}. Although at present
no microscopic mechanism is known to give $u_{2}\left(0\right)<0$
\cite{zlatko}, this could be another possibility to explain the existence
of nematic order without magnetic order in FeSe. Such an iCDW phase
could be detected
via its
time-reversal symmetry breaking, which would be manifested in, e.g.,
$\mu$SR measurements.
The phase diagram under pressure can be explained by  assuming
 that
  under pressure $u_{2}$ would change
sign and SDW would become the leading instability.

\textit{Summary}~~~ In summary, we propose a natural extension
of the Ising-nematic scenario to explain the puzzling nematic state
observed in FeSe. Our scenario relies on the smallness of $E_{F}$
and explains the onset of nematic order far from magnetism due to
the near degeneracy between the SDW channel and an iCDW charge-current
density wave channel. This near-degeneracy could result in the nucleation
of local iCDW order in the presence of point-like impurities, which
favor iCDW against SDW order \cite{Schmalian15}. While these fluctuations
cooperate with magnetic ones to break the tetragonal symmetry, they
compete for long-range order and reduce both $T_{N}$ and the magnetic
correlation length at the onset of nematic order. We argue that this
Ising-nematic scenario can also explain the observed non-monotonic
dependence of the nematic transition temperature $T_{s}$ upon pressure.

We thank A. Boehmer, I. Fisher, P. Hirschfeld, U. Karahasanovic, J.
Kang, S. Kivelson, I. Mazin, C. Meingast, R. Valenti, for useful discussions.
This work was supported by the Office of Basic Energy Sciences U.
S. Department of Energy under awards DE-FG02-ER46900 (AVC) and DE-SC0012336
(RMF) and the Deutsche Forschungsgemeinschaft through DFG-SPP 1458
\emph{Hochtemperatursupraleitung in Eisenpniktiden }(JS).


\begin{thebibliography}{10}
\bibitem{review} D. C. Johnston, Adv. Phys., \textbf{59}, 803 (2010);
D.N. Basov and A.V. Chubukov, Nature Physics \textbf{7}, 241 (2011);
J. Paglione and R. L. Greene, Nature Phys. \textbf{6}, 645 (2010);
P. C. Canfield and S. L. Bud'ko, Annu. Rev. Cond. Mat. Phys. \textbf{1},
27 (2010); H. H. Wen and S. Li, Annu. Rev. Cond. Mat. Phys. \textbf{2},
121 (2011); P. Dai, J. Hu, and E. Dagotto, Nature Phys. \textbf{8},
709 (2012).

\bibitem{Fernandes14} R. M. Fernandes, A. V. Chubukov, and J. Schmalian,
Nature Phys. \textbf{10}, 97 (2014).

\bibitem{Xu08} C. Xu, M. Mueller, and S. Sachdev, Phys. Rev. B \textbf{78},
020501(R) (2008).

\bibitem{Fang08} C. Fang, H. Yao, W.-F. Tsai, J.P. Hu, and S. A.
Kivelson, Phys. Rev. B \textbf{77} 224509 (2008).

\bibitem{Si11} E. Abrahams and Q. Si, J. Phys.: Condens. Matter \textbf{23},
223201 (2011).

\bibitem{igor_m}  M. D. Johannes and I. I. Mazin, Nature Phys. \textbf{5},
141 (2009).

\bibitem{Batista11}  Y. Kamiya, N. Kawashima, and C. D. Batista,
Phys. Rev. B \textbf{84}, 214429 (2011).

\bibitem{Lorenzana11} M. Capati, M. Grilli, and J. Lorenzana, Phys.
Rev. B \textbf{84}, 214520 (2011).

\bibitem{Brydon11} P. M. R. Brydon, J. Schmiedt, and C. Timm, Phys.
Rev. B \textbf{84}, 214510 (2011).

\bibitem{Fernandes12} R. M. Fernandes, A. V. Chubukov, J. Knolle,
I. Eremin and J. Schmalian, Phys. Rev. B \textbf{85}, 024534 (2012).

\bibitem{Dagotto13} S. Liang, A. Moreo, and E. Dagotto, Phys. Rev.
Lett. \textbf{111}, 047004 (2013).

\bibitem{Yamase15} H. Yamase and R. Zeyher, arXiv:1503.07646.

\bibitem{Fernandes13} R. M. Fernandes, A. E. B�hmer, C. Meingast,
and J. Schmalian, Phys. Rev. Lett. \textbf{111}, 137001 (2013).

\bibitem{Blomberg13} E. C. Blomberg, M. A. Tanatar, R. M. Fernandes,
I. I. Mazin, B. Shen, H.-H. Wen, M. D. Johannes, J. Schmalian, and
R. Prozorov, Nature Comm. \textbf{4}, 1914 (2013).

\bibitem{Meingast_FeSe} A.\LyXThinSpace E. B�hmer, T. Arai, F. Hardy,
T. Hattori, T. Iye, T. Wolf, H.\LyXThinSpace v. L�hneysen, K. Ishida,
and C. Meingast Phys. Rev. Lett. \textbf{114}, 027001 (2015).

\bibitem{Nakayama_ARPES_14} K. Nakayama, Y. Miyata, G.\LyXThinSpace N.
Phan, T. Sato, Y. Tanabe, T. Urata, K. Tanigaki, and T. Takahashi
Phys. Rev. Lett. \textbf{113}, 237001 (2014).

\bibitem{Ding_ARPES_15} P. Zhang, T. Qian, P. Richard, X. P. Wang,
H. Miao, B. Q. Lv, B. B. Fu, T. Wolf, C. Meingast, X. X. Wu, Z. Q.
Wang, J. P. Hu, and H. Ding, arXiv:1503.01390.

\bibitem{ZXShen_ARPES_15} Y. Zhang, M. Yi, Z.-K. Liu, W. Li, J. J.
Lee, R. G. Moore, M. Hashimoto, N. Masamichi, H. Eisaki, S. -K. Mo,
Z. Hussain, T. P. Devereaux, Z.-X. Shen, and D. H. Lu, arXiv:1503.01556.

\bibitem{Coldea15} M. D. Watson, T. K. Kim, A. A. Haghighirad, N.
R. Davies, A. McCollam, A. Narayanan, S. F. Blake, Y. L. Chen, S.
Ghannadzadeh, A. J. Schofield, M. Hoesch, C. Meingast, T. Wolf, and
A. I. Coldea, arXiv:1502.02917.

\bibitem{INS_FeSe_1} M. C. Rahn, R. A. Ewings, S. J. Sedlmaier, S.
J. Clarke, and A. T. Boothroyd, arXiv:1502.03838.

\bibitem{INS_FeSe_2} Q. Wang, Y. Shen, B. Pan, Y. Hao, M. Ma, F.
Zhou, P. Steffens, K. Schmalzl, T. R. Forrest, M. Abdel-Hafiez, D.
A. Chareev, A. N. Vasiliev, P. Bourges, Y. Sidis, H. Cao, and J. Zhao,
arXiv:1502.07544.

\bibitem{McQueen09} T. M. McQueen \emph{et al}, Phys. Rev. Lett.
\textbf{103}, 057002 (2009).

\bibitem{Imai09} T. Imai, K. Ahilan, F.\LyXThinSpace L. Ning, T.\LyXThinSpace M.
McQueen, and R.\LyXThinSpace J. Cava, Phys. Rev. Lett. \textbf{102},
177005 (2009).

\bibitem{Buchner_FeSe} S.-H. Baek, D. V. Efremov, J. M. Ok, J. S.
Kim, J. van den Brink, and B. B�chner, Nat. Mater. \textbf{14}, 210
(2014).

\bibitem{w_ku10} C. C. Lee, W. G. Yin, and W. Ku, Phys. Rev. Lett.
\textbf{103}, 267001 (2009)

\bibitem{devereaux10} C.-C. Chen, J. Maciejko, A. P. Sorini, B. Moritz,
R. R. P. Singh, and T. P. Devereaux, Phys. Rev. B \textbf{82}, 100504
(2010)

\bibitem{Phillips10} W. Lv, F. Kr�ger, and P. Phillips, Phys. Rev.
B \textbf{82}, 045125 (2010).

\bibitem{Phillips12} W.-C. Lee and P. W. Phillips, Phys. Rev. B \textbf{86},
245113 (2012).

\bibitem{Kontani12} S. Onari H. and Kontani, Phys. Rev. Lett. \textbf{109},
137001 (2012).

\bibitem{Kivelson_Lee_15} F. Wang, S. Kivelson, and D.-H. Lee, arXiv:1501.00844.

\bibitem{Si15} R. Yu and Q. Si, arXiv:1501.05926.

\bibitem{Mazin15} J. K. Glasbrenner, I. I. Mazin, H. O. Jeschke,
P. J. Hirschfeld, and R. Valenti, arXiv:1501.04946.

\bibitem{FeSe_dHvA} T. Terashima \emph{et al}, Phys. Rev. B \textbf{90},
144517 (2014).

\bibitem{zlatko} J. Kang and Z. Tesanovic, Phys. Rev. B \textbf{83},
020505 (2011).

\bibitem{FeSe_pressure1} M. Bendele, A. Ichsanow, Y. Pashkevich,
L. Keller, T. Strassle, A. Gusev, E. Pomjakushina, K. Conder, R. Khasanov,
and H. Keller, Phys. Rev. B \textbf{85}, 064517 (2012).

\bibitem{FeSe_pressure2} T. Terashima, N. Kikugawa, S. Kasahara,
T. Watashige, T. Shibauchi, Y. Matsuda, T. Wolf, A. E. B�hmer, F.
Hardy, C. Meingast, H. v. L�hneysen, and S. Uji, arXiv:1502.03548.

\bibitem{Eremin10} I. Eremin and A. V. Chubukov, Phys. Rev. B \textbf{81},
024511 (2010).

\bibitem{GraserSDDeg} S. Graser, T. A. Maier, P. J. Hirschfeld, and
D. J. Scalapino, New J. Phys. \textbf{11}, 025016 (2009).

\bibitem{comm} The
low-energy excitations in the band basis come from three orbitals
-- $d_{xz}$, $d_{yz}$, and $d_{xy}$. The two hole pockets are composed
predominantly of $d_{xz}$ and $d_{yz}$ orbitals, and the two electron
pockets of $d_{xy}$ and $d_{xz}$ orbitals ($(0,\pi)$ pocket) or
$d_{xy}$ and $d_{yz}$ orbitals ($(\pi,0)$ pocket).

\bibitem{Valenzuela14} \textit{\emph{L. Fanfarillo, A. Cortijo, and
B. Valenzuela, arxiv:1410.8488.}}

\bibitem{Chubukov_RG} A. V. Chubukov, D. Efremov, and I. Eremin,
Phys. Rev. B \textbf{78}, 134512 (2008); A. V. Chubukov, Physica C
\textbf{469}, 640 (2009); S. Maiti and A. V. Chubukov, Phys. Rev.
B \textbf{82}, 214515 (2010).


\bibitem{Vafek_Fernandes} R. M. Fernandes and O. Vafek, Phys. Rev.
B \textbf{90}, 214514 (2014); V. Cvetkovic and O. Vafek, Phys. Rev.
B 88, 134510 (2013).

\bibitem{Schmalian15} M. Hoyer, M. S. Scheurer, S. V. Syzranov, and
J. Schmalian, Phys. Rev. B \textbf{91}, 054501 (2015).\end{thebibliography}
\end{document}